\documentclass[runningheads]{llncs}
\usepackage{amsmath,graphicx,todonotes}
\usepackage{multirow,stfloats}
\usepackage{verbatim}

\usepackage{amsfonts}
\usepackage{cite}
\usepackage{longtable}
\usepackage{color}
\usepackage{enumitem}
\usepackage{multirow}
\usepackage{graphicx}
 \usepackage[misc]{ifsym}

\begin{document}
\title{Semantic Modeling of Textual Relationships in Cross-Modal Retrieval}
%
%
\author{Jing Yu\inst{1}
\and
Chenghao Yang\inst{2}
\and
Zengchang Qin(\Letter)\inst{2}
\and
Zhuoqian Yang\inst{2}
\and
Yue Hu\inst{1}
\and
Weifeng Zhang\inst{3}
}
\authorrunning{Jing Yu et al.}
%
\institute{Institute of Information Engineering, Chinese Academy of Sciences, China 
\and
Intelligent Computing and Machine Learning Lab, Beihang University, China
\and
College of Mathematics, Physics and Information Engineering, Jiaxing University, China \\
\email{yujing02@iie.ac.cn, \{zcqin, alanyang, yzhq97\}@buaa.edu.cn， weifeng.zhang@gmail.com}
}
\maketitle              
\begin{abstract}
Feature modeling of different modalities is a basic problem in current research of cross-modal information retrieval. 
Existing models typically project  
texts and images into one embedding space, in which semantically similar information will have a shorter distance.
Semantic modeling of textural relationships is 
notoriously difficult. 
In this paper, we propose an approach to model texts using a featured graph by integrating multi-view textual relationships including semantic relations, statistical co-occurrence, and prior relations in knowledge base. 
A dual-path neural network is adopted to learn multi-modal representations of information and cross-modal similarity measure jointly. 
We use a Graph Convolutional Network (GCN) for generating relation-aware text representations, and use a Convolutional Neural Network (CNN) with non-linearities for image representations. The cross-modal similarity measure is learned by distance metric learning.
Experimental results show that, by leveraging the rich relational semantics in texts, our model can outperform the state-of-the-art models by 3.4\% and 6.3\% on accuracy on two benchmark datasets. 

\end{abstract}
\section{Introduction}
\label{sec:intro}
Cross-modal information retrieval (CMIR), which enables queries from one modality to retrieve information in another, plays an increasingly important role in intelligent searching and recommendation systems. 
A typical solution of CMIR is to project features from different modalities into one common semantic space in order to  measure  cross-modal similarity directly. Therefore, feature representation is fundamental for CMIR research and has great influence on the retrieval performance.
Recently, Deep Neural Networks (DNN) achieve superior advances in cross-modal retrieval \cite{yu2018modeling,Lee2018Stacked}. For text-image retrieval, 
much effort has been devoted to vector-space models, such as the CNN-LSTM network \cite{Li2017Identity}, to represent multimodal data as ``flat'' features for both irregular-structured text data and grid-structured image data.
For image data, CNN can effectively extract hierarchies of visual feature vectors. However, for text data, the  ``flat'' features 
are seriously limited by their inability to capture complex structures hidden in texts \cite{yu2018modeling,Qin2016Topic} -- there are many implicit and explicit textual relations that characterize syntactic rules in text modeling \cite{Jiang2010Text}. Nevertheless, the possibility of infusing prior facts or relations (e.g., from a knowledge graph) into deep textual models is excluded by the great difficulty it imposes.

\begin{figure}[tp]
\centering
\includegraphics[width=0.65\textwidth, height=6cm]{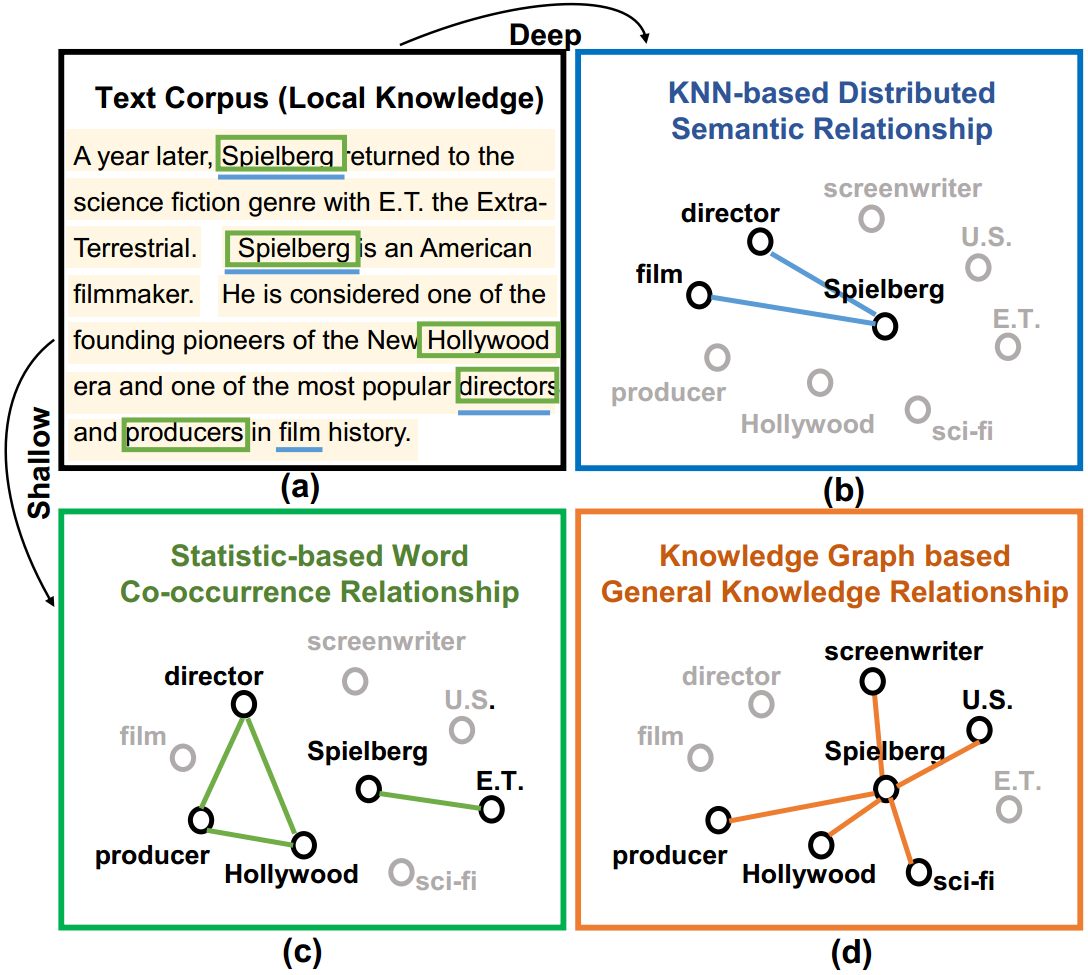}
\vspace*{-0.25cm}
\caption{(a) The original text and three kinds of textual relationships: (b) distributed semantic relationship in the embedding space, (c) word co-occurrence relationship and (d) general knowledge relationship defined by a knowledge graph.
}
\label{fig:3graph}
\end{figure}

Early works attempt to learn shallow statistical relationships, such as co-occurrence \cite{Rousseau2013graph} or location \cite{Mihalcea2004textrank}. 
Later on, semantic relationship based on syntactic analysis \cite{Jiang2010Text} or semantic rules between conceptual terms are explored. Besides, semantic relationship derived from knowledge graphs (e.g., Wikidata \cite{vrandevcic2014wikidata})  has attracted increasing attention. 
A most recent work \cite{yu2018modeling} models text as featured graphs with semantic relations. 
However, the performance of this practice heavily relies on the generalization ability of the word embeddings \cite{mikolov2013efficient}. 
It also 
fails to incorporate general human knowledge and other textual relations. 
To illustrate the above point, a text modeled by different types of relationships is shown in Fig.\ref{fig:3graph}.
It can be observed in the KNN graph (Fig.\ref{fig:3graph}-b) that \emph{Spielberg} is located relatively far away from \emph{Hollywood} as compared to the way \emph{director} is to \emph{film}, whereas in the common sense knowledge graph given in (Fig. \ref{fig:3graph}-d), these two words are closely related to each other as they should be.
Fig.\ref{fig:3graph}-c shows the less-frequent subject-predicate relation pattern (e.g. \emph{Spielberg} and \emph{E.T.}) which is absent in the KNN-based graph. Consequently, a more sophisticated model should correlate \emph{Spielberg} with all the following words \emph{\{director, film, E.T., Hollywood, producer, sci-fi, screenwriter, U.S.\}}. 
The above analysis indicates that graph construction can be improved by fusing different types of textual relations, which is the underlying motivation of this work.

In this paper, we propose a GCN-CNN architecture to learn textual and visual features for similarity matching. 
The novelty is on the in-depth study of textual relationship modeling for enhancing the successive correlation learning.
The key idea is to explore the effects of multi-view relationships and propose a graph-based integration model to combine complementary information from different given relationships. 
Specifically, besides semantic and statistic relations, we also exploit fusion with the relational knowledge bases (i.e., Wikidata \cite{vrandevcic2014wikidata}) for acquiring common sense about entities and their semantic relations, thus resulting in a knowledge-driven model. 
TensorFlow implementation of the model is available at \url{https://github.com/yzhq97/SCKR}.

\section{Methodology}
\label{sec:relation}

\begin{figure}[ht]
\centering
\includegraphics[width = 1.0 \textwidth]{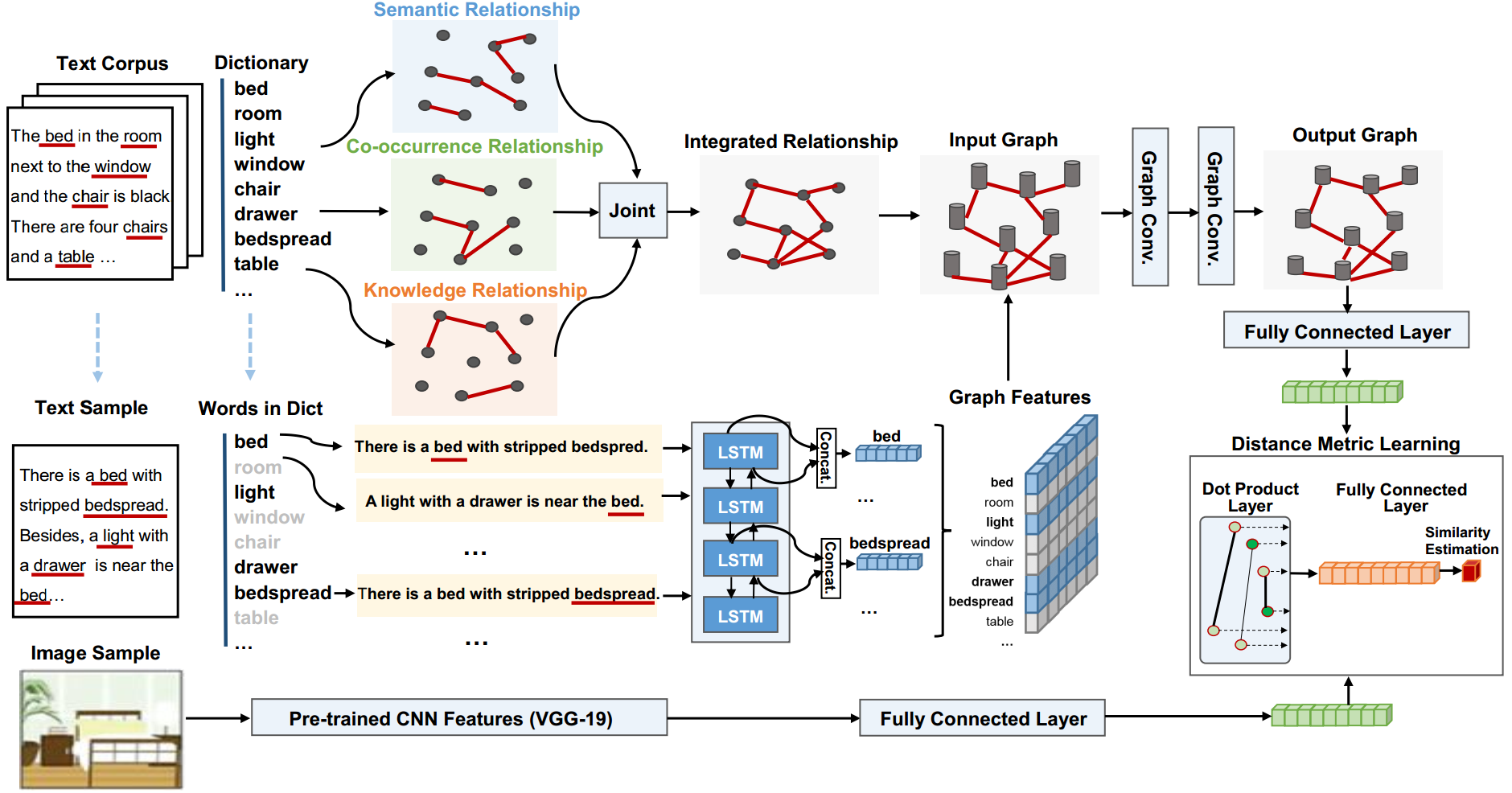}
\caption{The schematic illustration of our proposed framework for cross-modal retrieval.}
\label{fig:framework}
\end{figure}

In this paper, a dual-path neural network (as shown Fig.  \ref{fig:framework}) is proposed to learn multimodal features and cross-modal similarity in an end-to-end mode. It mainly consists of three parts:  (1) \textit{Text Modeling} (top in Fig.  \ref{fig:framework}): each text is represented by a featured graph by combining multi-view  relationships, that is also the key idea and will be elaborated later. Graph construction is performed off-line and the graph structure is identical for all the texts in the dataset. Then we adopt Graph Convolutional Network (GCN) \cite{Deff:fast}, containing two layers of convolution modules, to progressively enhance the textual representations over the constructed graph. The last FC layer projects the text features to the common semantic space; (2) \textit{Image Modeling} (bottom in Fig.  \ref{fig:framework}): we use pre-trained Convolutional Neural Network (CNN), i.e., VGGNet \cite{simonyan2015very}, for visual feature learning.
Similar to text modeling, the last FC layer is fine-tuned to project visual features to the same semantic space as the text; (3) \textit{Distance Metric Learning} (right in Fig.  \ref{fig:framework}): the similarity between textual and visual features are measured via distance metric learning. An inner product layer is used to combine these two kinds of features, followed by a FC layer with a sigmoid activation to output the similarity scores. We use ranking-based pairwise loss function formalized in \cite{kumar2016learning} for training, which can maximize the similarity of positive text-image pairs and minimizes the similarity of negative ones.

\subsection{Fine-grained Textual Relationship}
\label{subsec:graphconstruction}
In this section, we introduce the construction of graph structure to represent each text. As is mentioned above, all the texts share the same graph.
Given the training texts, we extract all the nouns to form a dictionary and each noun corresponds to a vertex in the graph. The vertex set is denoted as $V$. Edges are the integration of the following relationships from different views.  \\
\textbf{Distributed Semantic Relationship (SR)}
Following the distributional hypothesis \cite{harris1954distributional}, words appear in similar context may share semantic relationship, which is critical for relation modeling. To model such semantic relationship, we 
build a semantic graph denoted as $G_{SR}=(V,E_{SR})$.  Each edge
$e_{ij(SR)}\in E_{SR}$ is defined as follows:
\begin{equation}
\small
e_{ij(\text{SR})}=\begin{cases}
1 & \text{ if } w_i\in N_k(w_j)\  \text{or} \  w_j\in N_k(w_i)\\
0 & \text{ otherwise }
\end{cases}
\end{equation}
where $N_k(\cdot )$ is the set of $k$-nearest neighbors computed by the cosine similarity between words using  \textit{word2vec} embedding and
$k$ is the neighbor numbers, which is set to 8 in our experimental studies. \\ 
\textbf{Word Co-occurrence Relationship (CR)}
Co-occurrence statistics have been widely used in many tasks such as keyword extraction \cite{matsuo2004keyword} and web search \cite{matsuo2006graph}. 
Although the appearance of word embeddings seems to eclipse this method, we argue that 
it can serve as effective backup information to capture infrequent but syntax-relevant relations. 
Each edge $e_{ij(CR)}\in E_{CR}$ in the graph $G_{CR}=(V,E_{CR})$ indicates that the words $w_i$ and $w_j$ co-occur at least $\epsilon$ times.
The CR model can be formulated as below:
\begin{equation}
\small
e_{ij(\text{CR})}=\begin{cases}
1 & \text{ if } Freq(w_i,w_j) \geq \epsilon\\
0 & \text{ otherwise }
\end{cases}
\end{equation}
where $Freq(w_i,w_j)$ denotes the frequency that $w_i$ and $w_j$ appear in the same sentence in the dataset, we define $\epsilon$ as the threshold to rule out noise, which aims to achieve better generalization ability and improve computation efficiency. We empirically set $\epsilon$ to be $5$.\\
\textbf{General Knowledge Relationship (KR)}
General knowledge can effectively support decision-making and inference by providing  high-level expert knowledge as complementary information to training corpus. 
However, it is not fully covered by task-specific text. 
In this paper, we utilize
the triples in Knowledge Graphs (KG), i.e. (Subject, Relation, Predicate), which well represent various relationships in human commonsense knowledge.
To incorporate such real-world relationships, we construct the graph $G_{KR}=(V,E_{KR})$ and each edge $e_{ij(KR)} \in E_{KR}$ is defined as below:
\begin{equation}
\small
e_{ij(\text{KR})}=\begin{cases}
1 & \text{ if } (w_i, relation(w_i,w_j), w_j) \in D \\
0 & \text{ otherwise }
\end{cases}
\end{equation}
where $D$ refers to a given knowledge graph. In this paper, we adopt wikidata \cite{vrandevcic2014wikidata} in our experiments. For simplification, we ignore the types of relations in KG and leave it for the future work.
\\
\textbf{Graph Integration}
Different textual relationships capture information from different perspectives. 
It is conceivable that the relationship integration will fuse  semantic information. We simply utilize the union operation to obtain multi-view relationships.
$G=(V, E)$, where the edge set $E$ satisfying:

\begin{equation}
\small
    E = E_{SR} \cup E_{CR} \cup E_{KR}
\end{equation}

\subsection{Graph Feature Extraction}
\label{subsec:featureExtraction}
Previous work \cite{yu2018modeling} adopts Bag-of-Words (BoW), i.e., the word frequency, as the feature of each word in the text. However, this kind of feature is not informative enough to capture the rich semantic information. In this paper, we propose a kind of context-aware features for word-level representations. We first pretrain a Bi-LSTM \cite{graves2005framewise} in the text parts of the training set to predict the corresponding category labels, then sum up the concatenated outputs of Bi-LSTM  of each word over every mention in the text to obtain the word representation. Such representation is context-relevant and can better incorporate the content-specific semantics in the text. From our experiment observation, our proposed context-aware graph features can achieve +2\% overall retrieval performance lift compared with traditional BoW features. Due to the space limitation, we omit the BoW experimental results and focus on our proposed Bi-LSTM features.


\section{Experimental Studies}
\label{sec:experiments}

\textbf{Datasets.}
In this section, we test our models on two benchmark datasets: Cross-Modal Places \cite{castrejon2016learning} (CMPlaces) and English Wikipedia \cite{rasiwasia2010new} (Eng-Wiki). 
CMPlaces is one of the largest cross-modal datasets providing weakly aligned data in five modalities
divided into 205 categories. 
We follow the way in \cite{yu2018modeling} for sample generation, resulting in 204,800 positive pairs and 204,800 negative pairs for training, 1,435 pairs for validation and 1,435 pairs for test.
Eng-Wiki is the most widely used dataset in literature. There are 2,866 image-text pairs divided into 10 categories. We generate 40,000 positive samples and 40,000 negative samples respectively from the given 2,173 pairs for training. The remaining 693 pairs are for test. 
We use MAP@100 to evaluate the performance.
The density for all models over two datasets is much less than 1\%, indicating that our models are not trivial dense matrix. 

\textbf{Implementation Details.} We randomly selected 204,800 positive samples and 204,800 negative samples for training. We set the dropout ratio 0.2 at the input of the last fully connected layer, learning rate 0.001 with an Adam optimization, and regularization weight 0.005. The parameters setting for loss function follows \cite{yu2018modeling}. In the final semantic mapping layers of both text path and image path, the reduced dimensions are set to 1,024 for both datasets. The Bi-LSTM model is pretrained on classification task on Eng-wiki and CMPlaces, respectively.

\begin{table}[tp]
\centering
\small
\caption{MAP score comparison on two benchmark
datasets.}
\begin{tabular}{c|c|c|c|c}
\hline
\label{table:exp}
Method & $Q_T$ & $Q_I$ & Avg.  & Dataset                    \\ \hline\hline
CCA \cite{rasiwasia2010new}    & 18.7    & 21.6         & 20.2      & \multirow{15}{*}{Eng-Wiki}  \\
SCM \cite{rasiwasia2010new}    & 23.4    & 27.6        & 25.5     &          \\
LCFS \cite{Wang2013Learning}   & 20.4    & 27.1         & 23.8    &          \\
LGCFL \cite{Kang2015Learning}  & 31.6    & 37.8        & 34.7     &          \\
GMLDA \cite{Sharma2012Generalized}  & 28.9    & 31.6        & 30.2     &          \\
GMMFA \cite{Sharma2012Generalized}  & 29.6    & 31.6        & 30.6     &          \\
AUSL \cite{Zhang2017Adaptively}   & 33.2    & 39.7        & 36.4     &          \\
JFSSL \cite{Wang2016Joint}  & 41.0    &\textbf{46.7} & 43.9     &          \\
GIN \cite{yu2018modeling}    & 76.7    & 45.3        & 61.0     &          \\
\cline{1-4}
SR [ours]     & 83.5    & 41.4        & 62.4     &                \\
SCR [ours]   & 84.3   & 42.6       & 63.4    &                            \\
SKR [ours]   & 83.9    & 42.0       & 62.9    &                            \\
SCKR [ours] & \textbf{84.9} & 44.0      & \textbf{64.4} &      \\ \hline\hline
BL-ShFinal \cite{castrejon2016learning}& 3.3    & 12.7        & 8.0     & \multirow{9}{*}{CMPlaces}          \\
Tune(Free) \cite{castrejon2016learning}& 5.2 & 18.1 & 11.7 &          \\
TuneStatReg \cite{castrejon2016learning}& 15.1 & 22.1 & 18.6 &          \\
GIN \cite{yu2018modeling}    & 19.3     & 16.1   & 17.7  &          \\
\cline{1-4}
SR [ours] & 18.6       & 15.8   & 17.2    &                \\
SCR [ours]   & 25.4     & 20.3       & 22.8   &                            \\
SKR [ours]             & 24.8      & 20.5  & 22.6 &                            \\
SCKR [ours] & \textbf{28.5} & \textbf{21.3}           & \textbf{24.9} &      \\ \hline
\end{tabular}
\end{table}

\textbf{Comparison with State-of-the-Art Methods.}
In the Eng-Wiki dataset, we compare our model to some state-of-the-art (SOTA) retrieval models, which are listed in Table \ref{table:exp}.
We observe that SCKR achieves the best performance on the average MAP scores and slightly inferior to JFSSL on the image query ($Q_I$), which confirms that our relation-aware model can bring an overall improvement over existing CMIR models. Especially, text query ($Q_T$) gains remarkable 8.2\% increase over the SOTA model GIN, which proves that our model leads to better representation and generalization ability for the text query.
In the large CMPlaces dataset,
compared with the previous SOTA models, SCKR also achieves $6.3\%$ improvement compared to TuneStatReg \cite{castrejon2016learning}.

\textbf{Ablation Study.}
In this section, we conduct ablation experiments to evaluate the influence of the components in our proposed SCKR model.  We compare SCKR model to three ablated versions, i.e., SR, SCR and SKR. 
The retrieval performance is also listed in Table \ref{table:exp}. Compared to SR, both SCR and SKR achieve a significant improvement on both datasets (i.e., +5\% on CMPlaces and +2\% on Eng-Wiki). It indicates that either co-occurrence or the commonsense knowledge could provide complementary information to the distributed semantic relationship modeling. 
By integrating all kinds of textual relationships (SCKR), we obtain further promotion on MAP scores, especially on the relation-rich CMPlaces dataset. It is because that SR, CR or KR alone focuses on different views of relationships and their integration could bring more informative connections to the relational graph, thus facilitating information reasoning. 

\begin{figure}[!t]
\centering
\includegraphics[width = 0.7\textwidth, height=7.5cm]{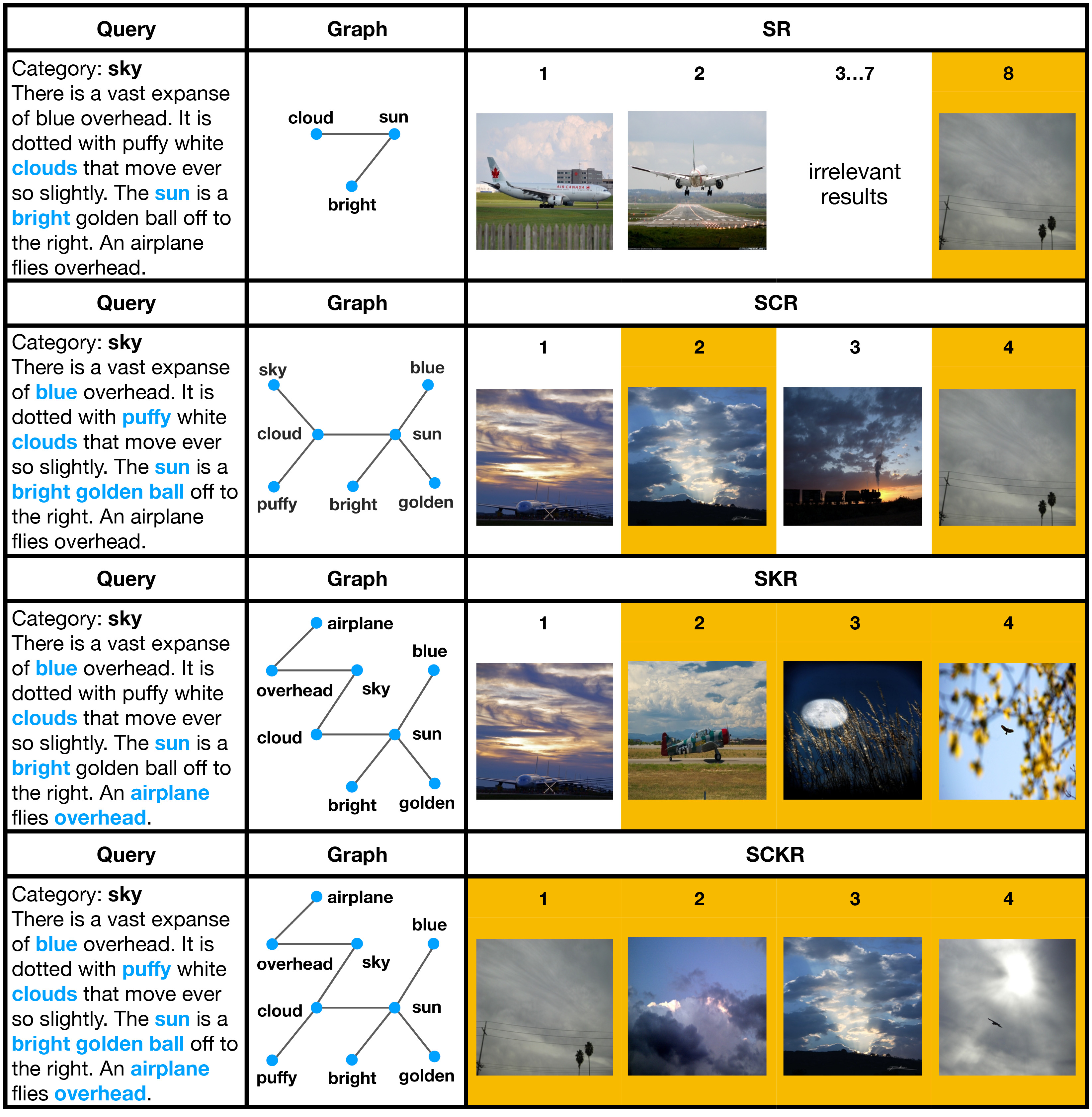}
\caption{
Some samples of text query results using four of our models on the CMPlaces dataset. The corresponding relation graphs are shown in the second column. 
The retrieval results are given in the third column.
}
\label{fig:cstq}
\end{figure}


\textbf{Qualitative Analysis.}
\label{subsec:qualitative}
Fig. \ref{fig:cstq} gives an example for the text-query task on SCKR and three baseline models. We show the corresponding relation graphs and the retrieved results. We observe that SR captures the least relationships and the results are far from satisfaction, which necessitates the exploration of the richer textual relationship. SCR can effectively emphasize the descriptive textual relationship (e.g. ``\textit{sun}-\textit{ball}" and ``\textit{sun}-\textit{bright}"), which is infrequent but informative for better understanding the content. Notice that, only SKR incorporates the relationship between ``\textit{overhead}'' and ``\textit{airplane}'' through ``\textit{sky}-\textit{overhead}-\textit{airplane}" inference path, which indicates that general knowledge is beneficial in relation inference and information propagation. The SCKR model leverages the advantages of different models and achieves the best performance.

%


\section{Conclusions}
\label{sec:conclusion}

In this paper, we proposed a graph-based neural model to integrate multi-view textual relationships, including the semantic relations, statistical co-occurrence, and pre-defined knowledge graph, for text modeling in CMIR tasks. The new model uses a GCN-CNN framework for feature learning and cross-modal semantic correlation modeling. Experimental results on both large-scale and widely-used benchmark datasets show that our model can significantly outperform the state-of-the-art models, especially for text queries. We achieve $\mathbf{3.4\%}$ and $\mathbf{6.3\%}$ improvement in accuracy comparing to state-of-the-art models respectively on Eng-Wiki and CMPlaces. 
In the future work, we can extend this model to other cross-modal areas such as automatic image captioning and video captioning.
\\
\textbf{Acknowledgement} This work is supported by the National Key Research and Development Program (Grant No.2017YFB0803301).

\bibliographystyle{splncs04}
\bibliography{references}

\begin{thebibliography}{10}
\providecommand{\url}[1]{\texttt{#1}}
\providecommand{\urlprefix}{URL }
\providecommand{\doi}[1]{https://doi.org/#1}

\bibitem{castrejon2016learning}
Castrejon, L., Aytar, Y., Vondrick, C., Pirsiavash, H., Torralba, A.: Learning
  aligned cross-modal representations from weakly aligned data. In: CVPR. IEEE
  (2016)

\bibitem{Deff:fast}
Defferrard, M., Bresson, X., Vandergheynst, P.: Convolutional neural networks
  on graphs with fast localized spectral filtering. In: NIPS. pp. 3837--3845
  (2016)

\bibitem{graves2005framewise}
Graves, A., Schmidhuber, J.: Framewise phoneme classification with
  bidirectional lstm and other neural network architectures. Neural Networks
  \textbf{18}(5-6),  602--610 (2005)

\bibitem{harris1954distributional}
Harris, Z.S.: Distributional structure. Word  \textbf{10}(2-3),  146--162
  (1954)

\bibitem{Jiang2010Text}
Jiang, C., Coenen, F., Sanderson, R., Zito, M.: Text classifcation using graph
  mining-based feature extraction. Knowledge-Based Systems  \textbf{23}(4),
  302--308 (2010)

\bibitem{Kang2015Learning}
Kang, C., Xiang, S., Liao, S., Xu, C., Pan, C.: Learning consistent feature
  representation for cross-modal multimedia retrieval. TMM  \textbf{17}(3),
  370--381 (2015)

\bibitem{kumar2016learning}
Kumar, V.B.G., Carneiro, G., Reid, I.: Learning local image descriptors with
  deep siamese and triplet convolutional networks by minimizing global loss
  functions. In: CVPR. p. 5385–5394 (2016)

\bibitem{Lee2018Stacked}
Lee, K.h., Chen, X., Hua, G., Hu, H., He, X.: Stacked cross attention for
  image-text matching. In: ECCV. p. arXiv:1803.08024 (2018)

\bibitem{Li2017Identity}
Li, S., Xiao, T., Li, H., Yang, W., Wang, X.: Identity-aware textual-visual
  matching with latent co-attention. In: ECCV. pp. 1908--1917 (2017)

\bibitem{matsuo2004keyword}
Matsuo, Y., Ishizuka, M.: Keyword extraction from a single document using word
  co-occurrence statistical information. International Journal on Artificial
  Intelligence Tools  \textbf{13}(01),  157--169 (2004)

\bibitem{matsuo2006graph}
Matsuo, Y., Sakaki, T., Uchiyama, K., Ishizuka, M.: Graph-based word clustering
  using a web search engine. In: EMNLP. pp. 542--550. Association for
  Computational Linguistics (2006)

\bibitem{Mihalcea2004textrank}
Mihalcea, R., Tarau, P.: Textrank: Bringing order into text. In: EMNLP. pp.
  404--411 (2004)

\bibitem{mikolov2013efficient}
Mikolov, T., Chen, K., Corrado, G., Dean, J.: Efficient estimation of word
  representations in vector space. arXiv preprint arXiv:1301.3781  (2013)

\bibitem{Qin2016Topic}
Qin, Z., Yu, J., Cong, Y., Wan, T.: Topic correlation model for cross-modal
  multimedia information retrieval. Pattern Analysis \& Applications
  \textbf{19}(4),  1007--1022 (2016)

\bibitem{rasiwasia2010new}
Rasiwasia, N., Costa~Pereira, J., Coviello, E., Doyle, G., Lanckriet, G.R.,
  Levy, R., Vasconcelos, N.: A new approach to cross-modal multimedia
  retrieval. In: ACMMM. pp. 251--260. ACM (2010)

\bibitem{Rousseau2013graph}
Rousseau, F., Vazirgiannis, M.: Graph-of-word and twidf: New approach to ad hoc
  ir. In: CIKM. pp. 59--68 (2013)

\bibitem{Sharma2012Generalized}
Sharma, A., Kumar, A., Daume, H., Jacobs, D.W.: Generalized multiview analysis:
  A discriminative latent space. In: CVPR. pp. 2160--2167 (2012)

\bibitem{simonyan2015very}
Simonyan, K., Zisserman, A.: Very deep convolutional networks for large-scale
  image recognition. In: ICLR (2015)

\bibitem{vrandevcic2014wikidata}
Vrande{\v{c}}i{\'c}, D., Kr{\"o}tzsch, M.: Wikidata: a free collaborative
  knowledgebase. Communications of the ACM  \textbf{57}(10),  78--85 (2014)

\bibitem{Wang2016Joint}
Wang, K., He, R., Wang, L., Wang, W., Tan, T.: Joint feature selection and
  subspace learning for cross-modal retrieval. PAMI  \textbf{38}(10),
  2010--2023 (2016)

\bibitem{Wang2013Learning}
Wang, K., He, R., Wang, W., Wang, L.: Learning coupled feature spaces for
  cross-modal matching. In: ICCV. pp. 2088--2095 (2013)

\bibitem{yu2018modeling}
Yu, J., Lu, Y., Qin, Z., Zhang, W., Liu, Y., Tan, J., Guo, L.: Modeling text
  with graph convolutional network for cross-modal information retrieval. In:
  PCM. pp. 223--234. Springer (2018)

\bibitem{Zhang2017Adaptively}
Zhang, L., Ma, B., He, J., Li, G., Huang, Q., Tian, Q.: Adaptively unified
  semi-supervised learning for cross-modal retrieval. In: IJCAI. pp. 3406--3412
  (2017)

\end{thebibliography}
%
%
%
%




\end{document}